\documentclass[sigconf]{acmart}

\AtBeginDocument{%
  \providecommand\BibTeX{{%
    \normalfont B\kern-0.5em{\scshape i\kern-0.25em b}\kern-0.8em\TeX}}}

\acmDOI{10.1145/1122445.1122456}

\usepackage{kotex}
\usepackage{multirow}
\usepackage{bbm}
\usepackage{physics}

\renewcommand\footnotetextcopyrightpermission[1]{}
\newcommand{\etal}{\textit{et al}. }
\newcommand{\ie}{\textit{i}.\textit{e}., }

\begin{document}


\title{False Negative Distillation and Contrastive Learning for Personalized Outfit Recommendation}


\author{Seongjae Kim, Jinseok Seol, Holim Lim, Sang-goo Lee}
    \affiliation{
        \institution{Seoul National University}
        \country{Republic of Korea}
    }
    \email{{sjkim, jamie, ihl7029, sglee}@europa.snu.ac.kr}







\begin{abstract}

    Personalized outfit recommendation has recently been in the spotlight with the rapid growth of the online fashion industry.
    However, recommending outfits has two significant challenges that should be addressed.
    The first challenge is that outfit recommendation often requires a complex and large model that utilizes visual information, incurring huge memory and time costs.
    One natural way to mitigate this problem is to compress such a cumbersome model with knowledge distillation (KD) techniques that leverage knowledge from a pretrained teacher model.
    However, it is hard to apply existing KD approaches in recommender systems (RS) to the outfit recommendation because they require the ranking of all possible outfits while the number of outfits grows exponentially to the number of consisting clothing items.
    Therefore, we propose a new KD framework for outfit recommendation, called False Negative Distillation (FND), which exploits \textit{false-negative} information from the teacher model while not requiring the ranking of all candidates.
    The second challenge is that the explosive number of outfit candidates amplifying the data sparsity problem, often leading to poor outfit representation.
    To tackle this issue, inspired by the recent success of contrastive learning (CL), we introduce a CL framework for outfit representation learning with two proposed data augmentation methods.
    Quantitative and qualitative experiments on outfit recommendation datasets demonstrate the effectiveness and soundness of our proposed methods.

\end{abstract}






\maketitle



\begin{figure}[t]
    \centering
    \includegraphics[width=\linewidth]{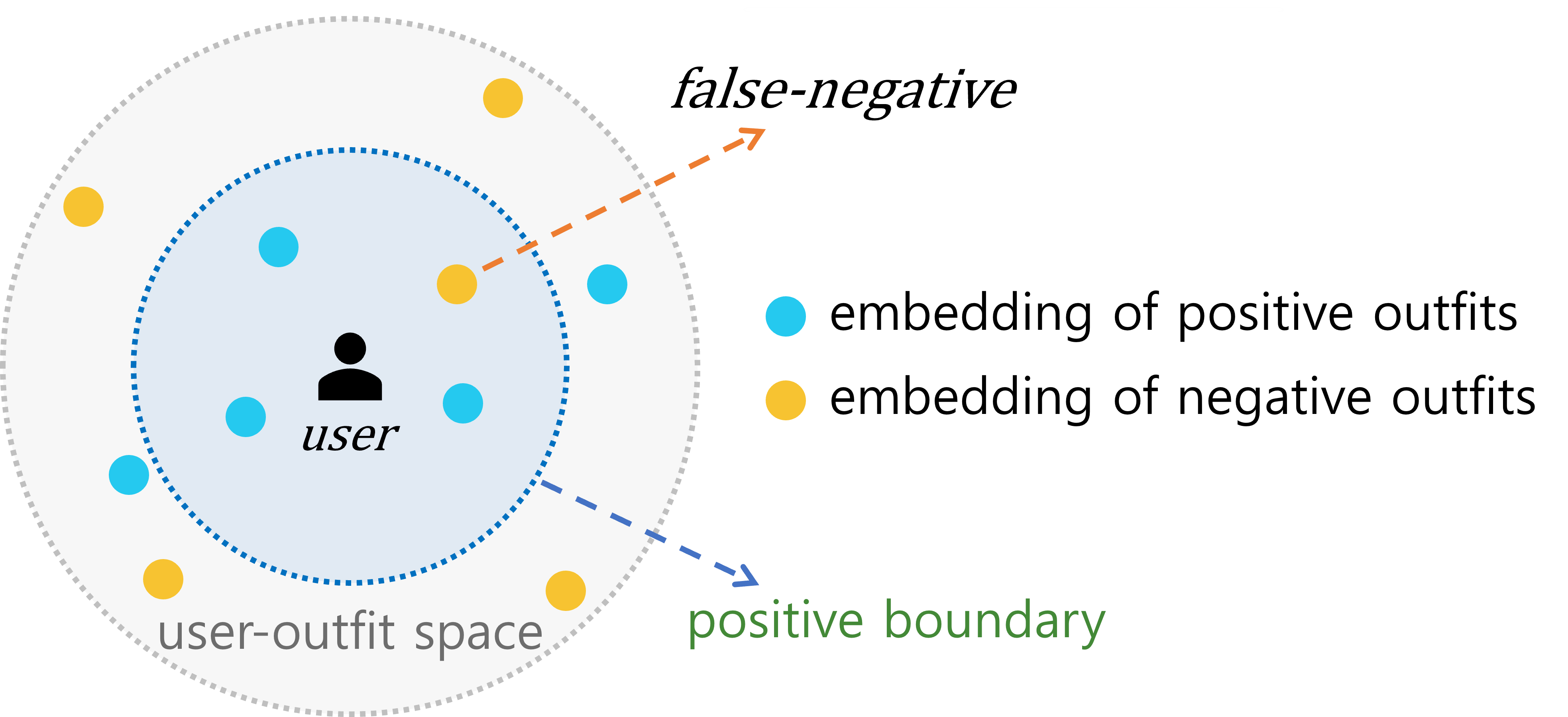}
    \caption{The user-outfit space is a representation vector space of the trained teacher model. We define the positive boundary as an average distance between a user and its positive outfits. We treat negative outfits inside the positive boundary as false-negative outfits.}
    \Description{A figure that shows the user-outfit space. In the space, there are user, positive outfits, negative outfits, and false-negative outfits. False-negative outfits are outfits that lie inside the positive boundary, where the model determines whether the outfit is preferable or not.}
    \label{fig:user-outfit space}
\end{figure}

\section{Introduction}


    Personalized outfit recommendation is the task of determining the preference of a user to an input outfit that consists of multiple clothing.
    It has recently attracted attention with the rapid growth of the online fashion industry, and several related studies \cite{OutfitNet, FHN, LPAE} have been conducted.
    However, despite the success of existing works, outfit recommendation has two significant challenges that should be addressed.
    First, recommending outfits often requires a complex and large model that involves the utilization of visual information (\ie images) \cite{LPAE}.
    Such a large model incurs high latency and memory costs during the inference phase, making it difficult to apply to real-time services \cite{DE-RRD}.
    The second challenge is that outfit recommendation inevitably suffers from the data sparsity problem because the possible pool of outfit data grows exponentially to the number of consisting clothing items \cite{OutfitNet}.
    The sparsity problem often leads to poor learning of outfit representation, which hinders achieving satisfactory recommendation performance \cite{ContrastiveSequential}.
    
    To address the first challenge stemming from a large model, one can employ knowledge distillation (KD) techniques that compress a model by transferring knowledge from a large teacher model to a small student model.
    Accordingly, one may try to apply existing studies \cite{RankingDistillation, CollaborativeDistillation, DE-RRD} of KD available in recommender systems (RS) to the outfit recommendation.
    However, existing methods leverage predicted ranking of all possible outfits from the teacher model, so they are not applicable in outfit recommendation tasks with explosively large pools.
    Therefore, we propose a novel KD framework named False Negative Distillation (FND) that does not require the ranking of all outfit candidates.
    Similar to most outfit recommendation studies \cite{OutfitNet, FHN, LPAE}, FND utilizes a ranking loss that pulls observed (positive) outfits to a user while pushing unobserved (negative) outfits.
    As illustrated in Fig. \ref{fig:user-outfit space}, FND claims that \textit{unobserved} is not the same as true-negative and assumes that negative outfits close enough to the user are false-negative outfits.
    We show through various experiments that FND is effective, and the assumption is reasonable.
    
\begin{figure}[t]
    \centering
    \includegraphics[width=\linewidth]{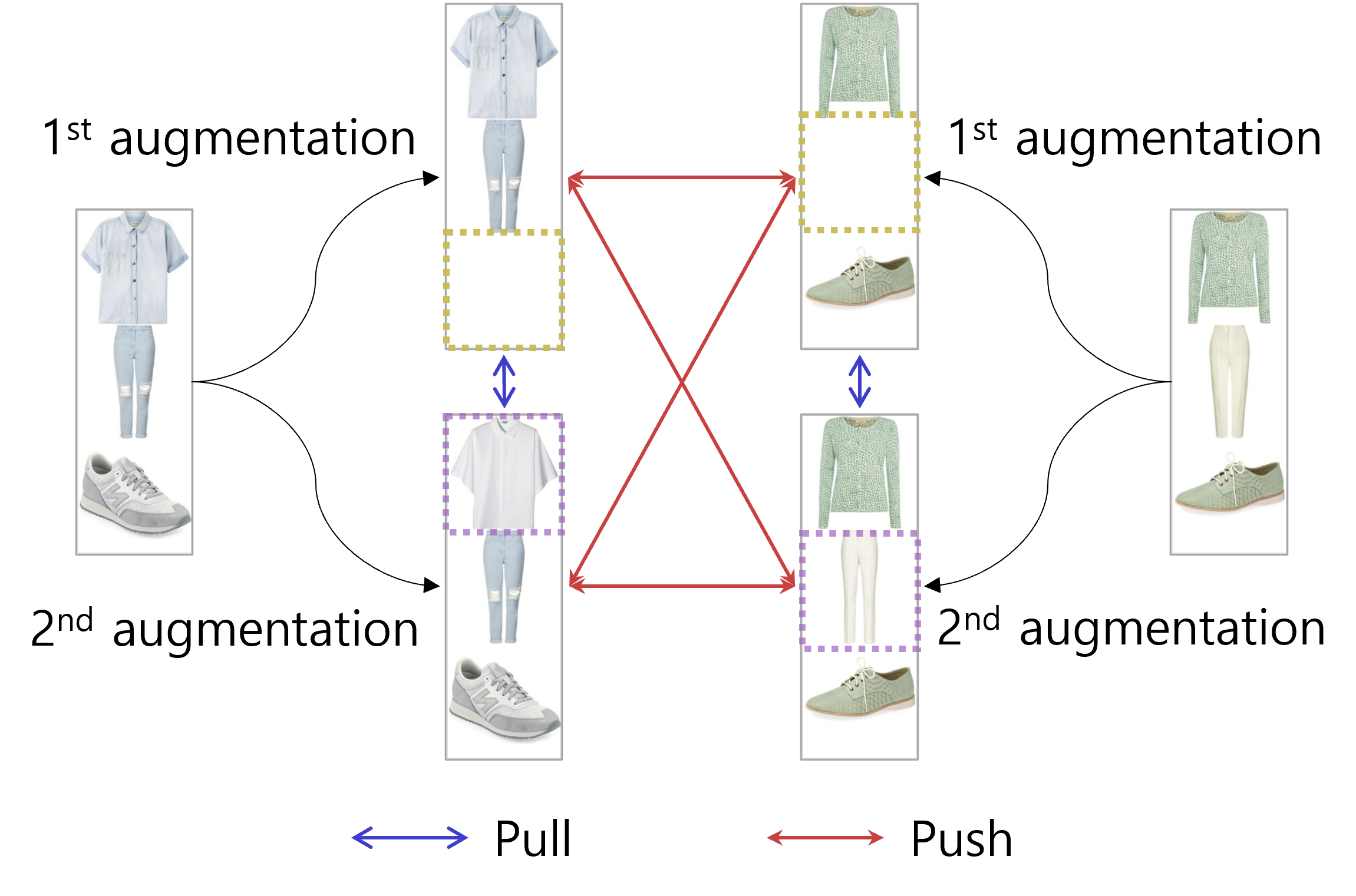}
    \caption{Example of contrastive learning for outfit recommendation. We randomly alter (\textit{erase}/\textit{replace}) one item from an outfit to generate two different views. In the case of \textit{replace}, it substitutes an item with a similar one using our pretrained autoencoder model. In this example, the first and the second augmentations are \textit{erase} and \textit{replace}, respectively. The left outfit erased the shoes and replaced the top. The right outfit erased the bottom and replaced the bottom.}
    \Description{A figure that shows examples of outfit augmentation methods and contrastive learning. Each outfit is augmented twice to generate two views, which should pull each other while views from the other outfits are pushed. Augmentation methods for outfits consist of erase and replace. Erase erases a random item from an outfit while replace randomly replaces an item. In the figure, there are two example outfits. One erased the shoes and replaced the top. The other erased the bottom and replaced the bottom.}
    \label{fig:contrastive learning}
\end{figure}

    The approach for the second challenge is to deal with the problem of poor outfit representation.
    As learning enhanced representation of entities is one of the core components to achieve high performance in deep learning \cite{Bengio2013}, numerous works \cite{Word2Vec, BERT, SSLImageRotation, SimCLR} from diverse domains accomplished this with self-supervised learning (SSL) techniques.
    There have been several studies \cite{DisentangledSequential, ReinforcementSequential, MutualSequential} utilizing SSL techniques in RS as well.
    Among them, more recent works \cite{ContrastiveSequential, ContrastiveItem, ContrastiveGraph} exploit contrastive learning (CL), especially SimCLR \cite{SimCLR}, which learns meaningful representations by pulling the differently augmented view of the same data while pushing the others in the batch.
    Nevertheless, leveraging CL in outfit recommendation is relatively unexplored.
    Hence, as illustrated in Fig. \ref{fig:contrastive learning}, we introduce an approach to make use of CL in outfit recommendation, along with two proposed data augmentation methods (\textit{erase}/\textit{replace}).
    
    To demonstrate the effectiveness of our proposed methods, we conduct extensive experiments on outfit recommendation datasets.
    We compare our approaches with state-of-the-art outfit recommendation methods with quantitative performance evaluations.
    In addition, we study the impact of hyper-parameters and the model size with various experiments.
    We use visualization to show that the intuitive assumption of our FND illustrated in Fig. \ref{fig:user-outfit space} is sound.
    We also experiment on the cold-start scenario where users have very few outfits interacted, and the trained model does not have any knowledge of those users.
    To make appropriate recommendations to cold starters, we introduce two practical strategies that do not require additional training of the model.
    
    
    Our main contributions can be summarized as follows:
    \begin{itemize}
        \item We propose a new knowledge distillation framework that can be utilized in outfit recommendation tasks without requiring the ranking of all outfit candidates in the system.
        \item We propose two novel outfit data augmentation methods to leverage contrastive learning in outfit recommendation.
        \item We introduce two practical strategies to deal with the cold-start problem.
        \item We demonstrate the effectiveness and soundness of our approaches with comprehensive experiments on fashion outfit recommendation datasets.
    \end{itemize}

\section{Related Work}


    \subsection{Outfit Recommendation}
    
        Based on whether the individual preference is neglected or not, existing outfit recommendation studies can be classified into two categories: non-personalized \cite{SiameseNet, Type-Aware, SCE-Net, Context-Aware, TC-GAE, SetNN, Bi-LSTM, MCAN} and personalized \cite{TensorFactorization, FHN, HFGN, OutfitNet, LPAE} outfit recommendation.
        Lu \etal \cite{FHN} used pairwise scores, and they employed the weighted hashing technique to tackle the efficiency problem.
        Lin \etal \cite{OutfitNet} utilized an attention mechanism to estimate the preference score, weighting items in an outfit differently.
        Lu \etal \cite{LPAE} exploited Set Transformer \cite{SetTransformer}, the state-of-the-art model for set-input problems, to capture the high-order interactions among fashion items.
        They also disentangled each user into multiple anchors to accommodate the variety of preferences.
        Note that methods based on graph neural networks \cite{Context-Aware, TC-GAE, HFGN} or predicting distribution over whole clothing items \cite{MCAN} require the test items to be in the training set.

    \subsection{Knowledge Distillation}
    
        Knowledge distillation is a model-agnostic compression strategy for generating efficient models.
        Since the early success of KD in image recognition \cite{KnowledgeDistillation, Fitnets}, KD has been widely accepted in other fields.
        In recommendation tasks, several works \cite{RankingDistillation, CollaborativeDistillation, DE-RRD} have employed KD.
        They rank all items with the teacher model and utilize the items of high rank when training the student model.
        Tang \etal \cite{RankingDistillation} considered top-$K$ items as false-negatives and differentiated their relative importance based on their rankings.
        Lee \etal \cite{CollaborativeDistillation} trained the student to mimic the predicted probabilities of the teacher on the sampled items of high rank.
        Kang \etal \cite{DE-RRD} achieved state-of-the-art performance by transferring both the prediction and latent knowledge of the teacher.
    
    
    
    \subsection{Contrastive Learning}
    
        Contrastive learning is a framework for obtaining high-quality representations to boost the performance of downstream tasks and was first introduced in computer vision \cite{SimCLR}.
        CL enhances representations by maximizing agreement between two differently augmented views of the same data.
        A few works \cite{ContrastiveSequential, ContrastiveItem, ContrastiveGraph} applied CL to RS, and they showed notable success.
        In sequential recommendation, Xie \etal \cite{ContrastiveSequential} used CL by applying three augmentation methods (crop/mask/reorder) to user interaction history.
        Yao \etal \cite{ContrastiveItem} focused on large-scale item recommendations and employed a two-stage augmentation consisting of masking and dropout.
        Liu \etal \cite{ContrastiveGraph} utilized CL for graph neural network based RS by randomly removing some edges.
    
    


\section{Approach}

    We recommend outfits to users based on their preference score.
    To compute the preference score, we use user embeddings and vector representations of outfits.
    Due to the set-like nature of fashion outfits, the representation model requires two conditions.
    First, the outfit representation should be invariant to the order of comprising fashion items.
    Second, the model should be able to process input outfits of any size.
    To this end, we borrow the architecture from LPAE \cite{LPAE} model, which uses Set Transformer \cite{SetTransformer} module designed to address these set-input problems.
    

    \subsection{Background: Computing the Preference Score to an Outfit}

\begin{figure}[t]
    \centering
    \includegraphics[width=\linewidth]{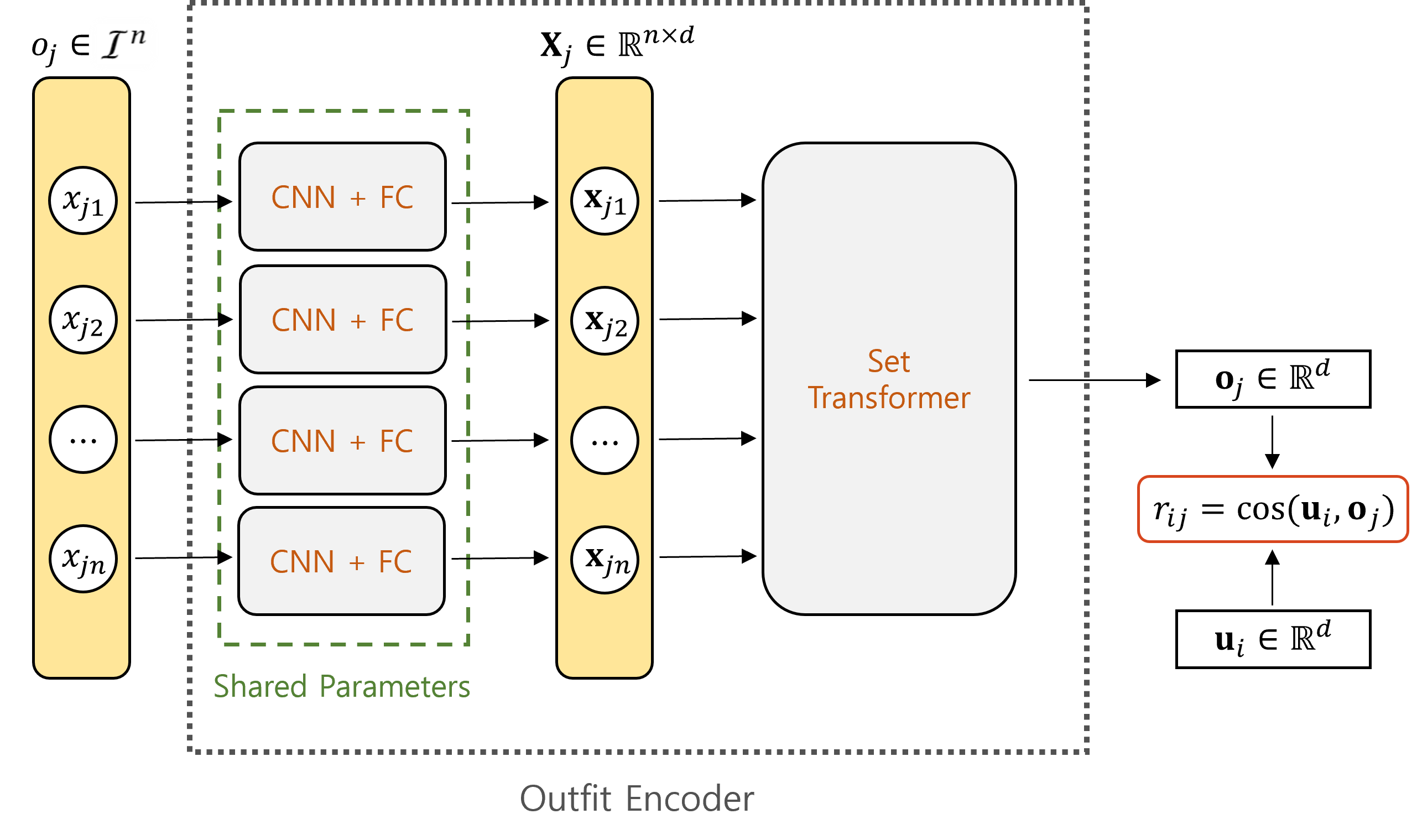}
    \caption{A brief architecture of computing the preference score of a user to an outfit. First, images comprising the outfit are transformed into item features by CNN with fully connected layers. After that, we get an outfit representation using Set Transformer. Finally, the cosine similarity between a user embedding and the outfit representation is the preference score of the user to the outfit.}
    \Description{An outfit is composed of multiple fashion images. First, the images are transformed into item features by CNN with fully connected layers. After that, we can get an outfit representation using Set Transformer. Finally, the cosine similarity between a user embedding and the outfit representation is the preference score of the user to the outfit.}
\label{fig:architecture}
\end{figure}

        As illustrated in Fig. \ref{fig:architecture}, an outfit $o_{j}$ with $n$ items is a tuple of fashion item images:
        \begin{math}
            o_{j} = (x_{j1}, x_{j2}, \cdots, x_{jn}) \in \mathcal{I}^{n}.
        \end{math}
        Let $f: \mathcal{I} \rightarrow \mathbb{R}^{d}$ be a Convolutional Neural Network (CNN) with fully connected layers that encodes $x_{jk}$ into an item feature vector $\vb{x}_{jk} = f(x_{jk})$, where $d$ is a feature dimension.
        Through \textit{Set Transformer} $T: \mathbb{R}^{\textit{n} \times \textit{d}} \rightarrow \mathbb{R}^{\textit{d}}$, we obtain an outfit representation $\vb{o}_{j} = T(\vb{X}_j)$ from item features
        \begin{math}
            \vb{X}_{j} = [
                \vb{x}_{j1} \,\,
                \vb{x}_{j2} \,\,
                \cdots \,\,
                \vb{x}_{jn}
            ]^{\top}.
        \end{math}
        Then, we compute the preference score with user embedding $\vb{u}_{i} \in \mathbb{R}^{d}$ for each user $u_{i}$.
        

        \subsubsection{Set Transformer}
        
            Having benefited from the \textit{attention} mechanism, Set Transformer can effectively reflect high-order interactions among items in an outfit.
            As Lee \etal \cite{SetTransformer} have proved, Set Transformer is an \textit{order-free} module that always produces the same output regardless of the sequence order.
            
            Attention is a map that gives the weighted sum of value vectors $\vb{V} \in \mathbb{R}^{n_{\text{v}} \times d_{\text{v}}}$ with the weights being determined by each query vector of $\vb{Q} \in \mathbb{R}^{n_{\text{q}} \times d_{\text{q}}}$ and key vectors $\vb{K} \in \mathbb{R}^{n_{\text{v}} \times d_{\text{q}}}$:
            \begin{equation}
            \label{eq:attn}
                \text{Attention}(\vb{Q}, \vb{K}, \vb{V}) =
                    \text{Softmax} (
                        \frac
                            {\vb{Q} \vb{K}^{\top}}
                            {\sqrt{d_{\text{q}}}}
                    )
                    \vb{V}
                .
            \end{equation}
            The \textit{multi-head attention} utilizes multiple attentions through concatenation to bear more potential relationships: for $h$ attention maps,
            \begin{equation}
                \label{eq:attn-head}
                \vb{A}_{i} = \text{Attention} (
                    \vb{Q} \vb{W}_{i}^{\text{Q}},
                    \vb{K} \vb{W}_{i}^{\text{K}},
                    \vb{V} \vb{W}_{i}^{\text{V}}
                ),
            \end{equation}
            \begin{equation}
                \label{eq:multihead}
                \text{MultiHead}(\vb{Q}, \vb{K}, \vb{V}) =
                    \text{Concat} (
                        \vb{A}_{1},
                        \vb{A}_{2},
                        \cdots,
                        \vb{A}_{h}
                    )
                    \vb{W}^{\text{M}}
                ,
            \end{equation}
            where
            \begin{math}
                \vb{W}_{i}^{\text{Q}}, \vb{W}_{i}^{\text{K}}
                    \in \mathbb{R}^{d_{\text{q}} \times d_{\text{q}}^{\text{M}}},
                \vb{W}_{i}^{\text{V}}
                    \in \mathbb{R}^{d_{\text{v}} \times d_{\text{v}}^{\text{M}}},
            \end{math}
            and
            \begin{math}
                \vb{W}^{\text{M}}
                    \in \mathbb{R}^{hd_{\text{v}}^{\text{M}} \times d_{\text{v}}}.
            \end{math}
            Following previous works \cite{SetTransformer, LPAE}, we use $d_{\text{q}}^{\text{M}} = d_{\text{q}} / h$ and $d_{\text{v}}^{\text{M}} = d_{\text{v}} / h$.
            
            To apply the attention mechanism to a set, Set Attention Block (SAB) uses \textit{self-attention} with residual terms: for item features $\vb{X}_j$,
            \begin{equation}
                \vb{H} = \text{LayerNorm}(
                    \vb{X}_j
                    +
                    \text{MultiHead}(
                        \vb{X}_j,
                        \vb{X}_j,
                        \vb{X}_j
                    )
                ),
            \end{equation}
            \begin{equation}
                \text{SAB} (\vb{X}_j) = 
                    \text{LayerNorm} (
                        \vb{H}
                        +
                        \sigma (\vb{H})
                    )
                ,
            \end{equation}
            where $\sigma$ is any row-wise feed-forward layer, and $\text{LayerNorm}(\cdot)$ is Layer Normalization \cite{LayerNorm}.
            Multiple SABs can be stacked to encode higher-order interactions among the items:
            \begin{equation}
                \vb{F} =
                    \text{SAB} (
                        \text{SAB} (
                            \vb{X}_{j}
                        )
                    )
                .
            \end{equation}
            The final outputs of the attention blocks are then aggregated by applying another multi-head attention on a learnable seed vector $\vb{s} \in \mathbb{R}^{d}$ as follows:
            \begin{equation}
                \vb{z} =
                    \text{LayerNorm} (
                        \vb{s} + \text{MultiHead}(\vb{s}, \vb{F}, \vb{F})
                    )
                ,
            \end{equation}
            \begin{equation}
                \vb{o}_{j} =
                    \text{LayerNorm} (
                        \vb{z} + \sigma (\vb{z})
                    )
                .
            \end{equation}
            The obtained $\vb{o}_{j} \in \mathbb{R}^{d}$ is a single compact vector representation of an outfit $o_{j}$, holding compatibility relationships among consisting fashion items.
            

        \subsubsection{Preference score prediction}
        
            Given a user $u_{i}$ and an outfit $o_{j}$, our model predicts the preference score of the user to the outfit as follows:
            \begin{equation}
                r_{ij} =
                    \cos(\vb{u}_{i}, \vb{o}_{j}) =
                    \frac{
                        \vb{u}_{i}^{\top}
                        \vb{o}_{j}
                    }{
                        \| \vb{u}_{i} \|
                        \| \vb{o}_{j} \|
                    }
                .
            \end{equation}
    

    \subsection{False Negative Distillation}
    
        Large models generally show relatively higher recommendation performance compared to their smaller counterparts.
        However, employing a small-sized model is necessary to reduce latency and memory costs during the inference phase.
        Therefore, we propose a novel knowledge distillation framework named False Negative Distillation (FND) that transfers \textit{false-negative} information extracted from a well-trained large teacher model to a small student model.
        As illustrated in Fig. \ref{fig:user-outfit space}, in the user-outfit space of a trained teacher model, we assume that negative (\ie unobserved) outfits close enough to the user are false-negative outfits.
        

        \subsubsection{Teacher model}
        \label{sec:teacher}
    
            Deep learning based recommendation models adopt learning to rank framework via deep metric learning in general.
            The goal is to maximize the ranking of positive outfits given a predicted preference score.
            Many existing works \cite{SCE-Net, OutfitNet, FHN}, including LPAE \cite{LPAE}, use triplet loss \cite{TripletLoss} or Bayesian personalized ranking (BPR) \cite{BPR} as an optimization objective.
            However, they often suffer from poor local optima, partially because the loss function employs only one negative outfit in each update \cite{Sohn2016}.
            To address this problem, our model utilizes $N$-pair loss \cite{Sohn2016}.
            Aided by the temperature-scaled cross-entropy, $N$-pair loss can take multiple negative outfits into account per positive outfit.
            Let the batch $\mathcal{B}$, size of $N$, be a set of pairs $(u_{i}, o_{j})$, indicating that the user $u_{i}$ prefers the outfit $o_{j}$.
            Each pair in the batch has a set of negative outfits $\{o_{j_{k}}\}_{k=1}^{\textit{N}}$, sampled in the training step.
            Note that the negative set mainly contains randomly generated outfits and even can include positive outfits of other users.
            Our objective for the teacher model is as follows:
            \begin{equation}
                \mathcal{L}_{N\text{-pair}} =
                    -
                    \frac{1}{N}
                    \sum_{(u_{i}, o_{j}) \in \mathcal{B}}
                        \log
                            \frac{
                                \exp(r_{ij} / \tau_{\text{FND}})
                            }{
                                \displaystyle
                                \exp(r_{ij} / \tau_{\text{FND}})
                                +
                                \sum_{k=1}^{N}
                                    \exp(r_{ij_{k}} / \tau_{\text{FND}})
                            }
                ,
            \end{equation}
            where $\tau_{\text{FND}} > 0$ is a temperature hyper-parameter.
            

        \subsubsection{Student model}

            Once the teacher model is trained, we optimize our student model with the help of additional false-negative information.
            The ordinary $N$-pair loss can be interpreted as ``pulling'' positive outfits to a user while ``pushing'' the negatives, similar to triplet loss with user-anchor.
            Whenever a given negative outfit is determined as a false-negative, we wish to pull it rather than pushing it.
            Concretely, we determine the false-negativeness $d_{ij_{k}}$ based on the difference between the preference score of a negative outfit and the average score of the positives: with $J_{i}^{+}$ a set of indices of positive outfits for a user $u_{i}$,
            \begin{equation}
                \hat{r}_{i} =
                    \frac
                        {1}
                        {|J_{i}^{+}|}
                    \sum_{j^{+} \in J_{i}^{+}}
                        \hat{r}_{ij^{+}}
                ,
            \end{equation}
            \begin{equation}
            \label{eq:alpha}
                d_{ij_{k}} =
                    \alpha (
                        \hat{r}_{i}
                        -
                        \hat{r}_{ij_{k}}
                    )
                ,
            \end{equation}
            where $\alpha > 0$ is a distillation scaling hyper-parameter and $\hat{r}_{ij}$ denotes the predicted score from the teacher.
            The sign of $d_{ij_{k}}$ determines whether the given negative outfit is false-negative or not, and the magnitude presents how much the negative should be pushed or pulled.
            The student model is trained through our proposed FND loss $\mathcal{L}_{\text{FND}}$ as follows:
            \begin{equation}
                \ell_{\text{FND}}(i, j) =
                    -
                    \log
                    \frac{
                        \exp(r_{ij} / \tau_{\text{FND}})
                    }{
                        \displaystyle
                        \exp(r_{ij} / \tau_{\text{FND}})
                        +
                        \sum_{k=1}^{N}
                            \exp(d_{ij_{k}} r_{ij_{k}} / \tau_{\text{FND}})
                    }
                ,
            \end{equation}
            \begin{equation}
                \mathcal{L}_{\text{FND}} =
                    \frac{1}{N}
                    \sum_{(u_{i}, o_{j}) \in \mathcal{B}}
                        \ell_{\text{FND}}(i, j)
                .
            \end{equation}

            Since the teacher model is frozen when training the student, $d_{ij_{k}}$ stays constant for each $i$ and $j_{k}$.
            Note that we use different validation sets when training the teacher model and the student model.
            Otherwise, the student model might learn information about the validation set via the teacher, which leads to overfitting.
            
            By examining the gradient of $\mathcal{L}_{\text{FND}}$, it can be shown that our objective function pulls negative outfits to the user whenever they are determined as false-negatives by $d_{ij_{k}}$.
            Suppose $o_{j_{k'}}$ is an element of the set of negative outfits $\{o_{j_{k}}\}_{k=1}^{N}$ for $(u_{i}, o_{j})$ in the training step.
            The gradient of $\ell_{\text{FND}}$ \textit{w.r.t} the preference score of $u_{i}$ to $o_{j_{k'}}$ is as follows:
            \begin{equation}
                p(r_{ij_{k'}}) =
                    \frac{
                        \exp(d_{ij_{k'}} r_{ij_{k'}} / \tau_{\text{FND}})
                    }{
                        \exp(r_{ij} / \tau_{\text{FND}})
                        +
                        \sum_{k=1}^{N}
                            \exp(d_{ij_{k}} r_{ij_{k}} / \tau_{\text{FND}})
                    }
                ,
            \end{equation}
            \begin{equation}
                \frac{
                    \partial \ell_{\text{FND}}(i, j)
                }{
                    \partial r_{ij_{k'}}
                } =
                    \frac
                        {d_{ij_{k'}}}
                        {\tau_{\text{FND}}}
                    p(r_{ij_{k'}})
                .
            \end{equation}
            The sign of the gradient is the same as $d_{ij_{k'}}$ since $\tau_{\text{FND}} > 0$ and $p(r_{ij_{k'}}) > 0$ hold.
            Hence, negative outfits closer than the average of positive outfits are pulled toward the user rather than pushed.
            

    \subsection{Contrastive Learning for Outfits}

        To obtain more enhanced outfit representations, we propose a novel approach to leverage SimCLR \cite{SimCLR} framework in outfit recommendation.
        Specifically, we suggest two data augmentation methods for outfits.

        SimCLR learns representations by maximizing the agreement between differently augmented views of the same outfit while pushing the others in the batch.
        Given the batch $\mathcal{B} = \{(u_{i^{(n)}}, o_{j^{(n)}})\}_{n=1}^{N}$, each outfit $o_{j^{(n)}}$ is augmented twice to create two different views $(\overline{o}_{j^{(2n-1)}}, \overline{o}_{j^{(2n)}})$, generating $2N$ augmented outfits $\{\overline{o}_{j^{(n)}}\}_{n=1}^{2N}$ in total.
        The agreement is measured by cosine similarity between each outfit representation: $s_{n, m} = \cos(g(\overline{\vb{o}}_{j^{(n)}}), g(\overline{\vb{o}}_{j^{(m)}}))$, where $g(\cdot)$ is a non-linear projection layer.
        The objective of contrastive learning is as follows:
        \begin{equation}
            \ell_{\text{CL}} (n, m) =
                -
                \log
                \frac{
                    \exp(s_{n,m} / \tau_{\text{CL}})
                }{
                    \sum_{t=1}^{2N}
                    \mathbbm{1}_{[t \ne n]}
                    \exp(s_{n,t} / \tau_{\text{CL}})
                }
            ,
        \end{equation}
        \begin{equation}
            \label{eq:cl}
            \mathcal{L}_{\text{CL}} =
                \frac
                    {1}
                    {2N}
                \sum_{n=1}^{N}
                    [
                        \ell_{\text{CL}}(2n-1, 2n)
                        +
                        \ell_{\text{CL}}(2n, 2n-1)
                    ]
            ,
        \end{equation}
        where $\tau_{\text{CL}} > 0$ is a temperature hyper-parameter.
        
        To exploit the objective in Eq. \ref{eq:cl}, we must define appropriate data augmentation methods which produce semantically similar outfits with an input outfit.
        As illustrated in Fig. \ref{fig:contrastive learning}, we suggest two augmentation methods suitable for outfit recommendation: \textit{erase} and \textit{replace}.
        Both augmentations randomly alter comprising items from an outfit while preserving the semantic context.
        We treat the augmentation set as a hyper-parameter and fix them at the beginning of the training.
        Note that if two identical augmentations are applied, we alter different items from the input outfit to obtain distinct views.
        

        \subsubsection{Erase}
        
            Randomly erasing components from the input is a common data augmentation method in diverse domains.
            In sequential recommendation, for example, Xie \etal \cite{ContrastiveSequential} randomly crop items from user interaction history.
            In natural language processing, Wu \etal \cite{CLEAR} erase or replace randomly selected words in a sentence.
            In computer vision, DeVries \etal \cite{Cutout} cut out contiguous sections of an input image, inspired by the object occlusion problem.
            When it comes to outfit recommendation, a subset of an outfit may imply or even determine the semantic information.
            Motivated by this, we randomly remove one item from the outfit to generate an augmented view.
        
        
        \subsubsection{Replace}
            
            Our proposed model computes preference scores based solely on visual information (\ie images), and this assumption is helpful if access to other metadata is limited.
            In this situation, the semantic information of an outfit is derived from the appearance of consisting items.
            Accordingly, we claim that the visual similarity of consisting items leads to the semantic similarity of an outfit.
            Based on the claim, we generate an augmented view by randomly replacing one item from the outfit with a visually similar item from the same category.
            To this end, we train a CNN autoencoder model and retrieve similar items through their latent features.


    \subsection{Final Objective: FND-CL}
    
        The proposed losses $\mathcal{L}_{\text{FND}}$ and $\mathcal{L}_{\text{CL}}$ can be used independently; hence we can take advantage of both methods.
        Therefore, our final objective is to minimize the weighted sum of both losses as follows:
        \begin{equation}
            \mathcal{L}_{\text{FND-CL}} =
                \mathcal{L}_{\text{FND}}
                +
                \lambda
                \mathcal{L}_{\text{CL}}
            ,
        \end{equation}
        where $\lambda$ is a loss weight hyper-parameter.


    \subsection{Profiling Cold Starters}


        In application services with recommender systems, new users can join the service even after the model is trained and deployed.
        Such users, or \textit{cold starters}, have relatively few interactions in general, and the deployed model does not have any prior knowledge of those users.
        In practice, fine-tuning the model for them is a time-consuming process; thus, cold starters might starve for the recommendation until the next iteration of deployment.
        Therefore, it is necessary to have an alternative recommendation method that exploits the already deployed model with no additional training.
        In personalized outfit recommendation, only a few works \cite{LPAE} handled the cold-start problem without fine-tuning the model.
        Here, we introduce two strategies to compute preference scores of cold starters, analogous to memory-based collaborative filtering.
        
        Given a cold starter $u_{c}$, we define a neighborhood $\mathcal{N}_{c}$ from the set of non-cold users $\mathcal{U}$ as follows: with a set of indices of positive outfits $J_{c}^{+}$ for $u_{c}$,
        \begin{equation}
            s_{ci} =
                \frac
                    {1}
                    {|J_{c}^{+}|}
                \sum_{j^{+} \in J_{c}^{+}}
                    r_{ij^{+}}
            ,
        \end{equation}
        \begin{equation}
            \mathcal{N}_{c} =
                \{
                    u_{i} \in \mathcal{U}
                \,|\,
                    s_{ci} > \delta
                    \vee
                    i = i_{c}^{*} 
                \}
            ,
        \end{equation}
        where $s_{ci}$ represents the asymmetric similarity from $u_{c}$ to $u_{i}$.
        $\delta$ is a similarity threshold, and $i_{c}^{*} = \text{argmax}_{i} s_{ci}$ denotes the index of the most similar user, which ensures at least one neighbor for each $u_{c}$.
        To compute the preference score $r_{cj}^{\text{cold}}$ of the cold starter $u_{c}$ to a given outfit $o_{j}$, we aggregate the preference scores of neighbors to the outfit.
        Here we use two aggregation strategies: \textit{Average} and \textit{Weighted Average}.


        \subsubsection{Average (avg)}

            A basic aggregation strategy is simply averaging the preference scores:
            \begin{equation}
                r_{cj}^{\text{cold}} :=
                    \frac
                        {1}
                        {|\mathcal{N}_{c}|}
                    \sum_{u_{i} \in \mathcal{N}_{c}}
                        r_{ij}
                .
            \end{equation}

        
        \subsubsection{Weighted Average (w-avg)}
    
            We further utilize the similarity between the cold starter and its neighbors as aggregation weights using cross-entropy with temperature:
            \begin{equation}
                \overline{s}_{ci} =
                    \frac
                        {\exp(s_{ci} / \tau_{\text{w-avg}})}
                        {
                            \sum_{u_{i'} \in \mathcal{N}_{c}}
                                \exp(s_{ci'} / \tau_{\text{w-avg}})
                        }
                ,
            \end{equation}
            \begin{equation}
                r_{cj}^{\text{cold}} :=
                    \sum_{u_{i} \in \mathcal{N}_{c}}
                        \overline{s}_{ci}
                        r_{ij}
                ,
            \end{equation}
            where $\tau_{\text{w-avg}} > 0$ is a temperature hyper-parameter, and note that $\sum_{u_{i} \in \mathcal{N}_{c}} \overline{s}_{ci} = 1$ holds.
            We considered other methods for deriving the aggregation weights $\overline{s}_{ci}$ from $s_{ci}$; however, the suggested method empirically showed the best and stable results, especially in terms of robustness to hyper-parameters.



\section{Experiment}


    \subsection{Experimental Design}
    
    
        \subsubsection{Datasets}

\begin{table}
    \caption{Dataset statistics.}
    \label{tab:dataset}
    \begin{tabular}{lrr}
        \toprule
        Dataset     & \# Outfits & \# Items \\
        \midrule
        Polyvore-630 & 162,945    & 199,537  \\
        Polyvore-519 & 106,806    & 178,481  \\
        Polyvore-53  & 13,707     & 26,727   \\
        Polyvore-32  & 6,565      & 18,656   \\
        \bottomrule
    \end{tabular}
\end{table}

            We use datasets collected from the Polyvore website: Polyvore-$U$ \cite{FHN}, where $U \in \{630, 519, 53, 32\}$ denotes the number of users.
            Polyvore-$U$ contains outfits posted by users, each consisting of three categories: top, bottom, shoes.
            Outfits in Polyvore-\{630, 53\} have a fixed number of items: one item for each category.
            Polyvore-\{519, 32\} include outfits with a variable number of items (\ie some outfits may have two tops).
            We use Polyvore-\{630, 519\} for most of the experiments and Polyvore-\{53, 32\} for cold starter tasks.
            Statistics of the datasets are provided in Table \ref{tab:dataset}.
            Following previous works \cite{FHN, LPAE}, we define user-posted outfits as positive outfits for each user and category-wise random mixtures of items as negative outfits.
            We also discuss the results of hard negative outfits (\ie random samples of positive outfits of other users) separately in Sec. \ref{sec:hard}.
            In the evaluation phase, we set the ratio between positive and negative outfits to 1:10 for each user.
            We split training, validation, and test sets to 9:2:2, and we further split the validation set into two halves, one for the teacher model and the other for the student model.
            As \cite{FHN} affirmed, there are no duplicate items between the training and the test sets for each user.
        
        
        \subsubsection{Evaluation metrics}
        
            We evaluate the ranking performance via Area Under the ROC curve (AUC) and Normalized Discounted Cumulative Gain (NDCG), similar to previous works \cite{FHN, LPAE}.
            For each user, we rank the test outfits by the predicted preference score of the model.
            We report the performance averaged over all users.
        
        
        \subsubsection{Considered methods}
        
            We compare our methods with the following state-of-the-art non-personalized \cite{Type-Aware, SCE-Net, Bi-LSTM} and personalized \cite{OutfitNet, FHN, LPAE} outfit recommendation models.
            Type-Aware \cite{Type-Aware} projects pairs of items onto the type-specific subspaces.
            Compatibility is then measured in these subspaces and learned through the triplet loss.
            SCE-Net \cite{SCE-Net} learns conditional embeddings and their weights using an attention mechanism.
            Each conditional embedding is implicitly encouraged to encode different semantic subspaces via the triplet loss.
            Bi-LSTM \cite{Bi-LSTM} considers an outfit as a sequence of items and uses a bidirectional LSTM to learn the compatibility.
            The model is trained by predicting the next and previous items in the sequence through cross-entropy loss.
            OutfitNet \cite{OutfitNet} consists of two stages to capture both general compatibility and personal taste.
            The objective of both stages is to maximize the difference between positive and negative scores, similar to BPR.
            FHN \cite{FHN} uses pairwise scores to compute outfit compatibility and personal preference simultaneously.
            We train FHN with BPR without the binarization step, following the previous work \cite{LPAE}.
            LPAE \cite{LPAE} includes two models LPAE-\textit{u} and LPAE-\textit{g}, which mainly handles the cold-start problem using multiple anchors for each user.
            Both models utilize BPR loss, and LPAE-\textit{g} has additional general anchors to model non-personalized compatibility.
            For a more fair comparison, we apply temperature scaling when using BPR or cross-entropy loss.
        
        
        \subsubsection{Implementation details}
        
            Similar to the previous work \cite{FHN}, we use \textit{AlexNet} \cite{AlexNet} pretrained on ImageNet \cite{Imagenet} as a backbone CNN.
            We define two versions of AlexNet to experiment the knowledge distillation.
            One is AlexNet-large, which is the original AlexNet.
            The other is AlexNet-small, a downsized version of AlexNet that all fully-connected layers are removed and with a global average pooling at the end \cite{GAP}.
            Only the teacher model uses AlexNet-large, and the others use AlexNet-small.
            We set the feature dimension $d = 128$ for all methods.
            For simplicity, we set $\delta$ to zero, the median of the possible range of cosine similarity.
            We use $\tau_{\text{FND}} = \tau_{\text{CL}} = 0.1$, $\tau_{\text{w-avg}} = 0.2$, $\lambda = 0.2$, and the number of heads $h = 8$.
            We set $\alpha$ to 1.25 for Polyvore-630 and 1.5 for Polyvore-519.
            When it comes to CL, the pair of augmentations are (\textit{erase}, \textit{replace}) for Polyvore-630 and (\textit{erase}, \textit{erase}) for Polyvore-519.
            SGD with momentum \cite{SGDwithMomentum} is used to train all methods, and the batch size is set to 32.
            For each method, we report the test performance with their optimal hyper-parameters searched via the validation set unless otherwise specified.
    
    
    \subsection{Performance Comparison}

\begin{table}
    \caption{Performance comparison of different methods on Polyvore datasets.}
    \label{tab:main}
    \begin{tabular}{lcccc}
        \toprule
                                     & \multicolumn{2}{c}{Polyvore-630} & \multicolumn{2}{c}{Polyvore-519}   \\
        \midrule
        Method                       & AUC             & NDCG           & AUC             & NDCG             \\
        \midrule
        \midrule
        Type-Aware \cite{Type-Aware} & 75.87           & 57.06          & 78.03           & 60.33            \\
        SCE-Net \cite{SCE-Net}       & 77.76           & 59.83          & 78.86           & 61.37            \\
        Bi-LSTM \cite{Bi-LSTM}       & 78.30           & 62.30          & 80.25           & 65.54            \\
        \midrule
        OutfitNet \cite{OutfitNet}   & 85.03           & 72.87          & 85.04           & 71.96            \\
        FHN \cite{FHN}               & 87.15           & 76.33          & 87.89           & 77.07            \\
        LPAE-\textit{u} \cite{LPAE}  & 87.82           & 77.49          & 89.19           & 79.23            \\
        LPAE-\textit{g} \cite{LPAE}  & 87.05           & 76.33          & 89.59           & 81.05            \\
        \midrule
        FND                          & 89.91           & 82.18          & 91.86           & 84.74            \\
        FND-CL                       & \textbf{90.28}  & \textbf{82.47} & \textbf{92.30}  & \textbf{85.51}   \\
        \bottomrule
    \end{tabular}
\end{table}

        As shown in Table \ref{tab:main}, our proposed FND outperforms baseline methods under all datasets and metrics.
        Furthermore, the performance of FND-CL shows the effectiveness of the outfit CL framework.
        
        Recall that LPAE-\textit{u} adopts multiple anchors for representing each user, and LPAE-\textit{g} further leverages non-personalized compatibility.
        Comparison with LPAE models shows that FND can effectively achieve improved performance without auxiliary parameters and structures.

    \subsection{Performance on Cold Starters}

\begin{table}
    \caption{Comparison of different methods on cold starters. For our FND and FND-CL, w/o and w/ ``(w)'' represent the \textit{avg} and \textit{w-avg} strategies, respectively. We use AUC as an evaluation metric.}
    \label{tab:cold}
    \begin{tabular}{lcccc}
        \toprule
                                     & \multicolumn{2}{c}{Polyvore-53} & \multicolumn{2}{c}{Polyvore-32}  \\
        \midrule
        \# Outfits                   & 1               & 5             & 1                & 5             \\
        \midrule
        \midrule
        Type-Aware \cite{Type-Aware} & 72.79          & 72.79          & 77.44           & 77.44          \\
        SCE-Net \cite{SCE-Net}       & 76.55          & 76.55          & 78.97           & 78.97          \\
        Bi-LSTM \cite{Bi-LSTM}       & 78.61          & 78.61          & 81.69           & 81.69          \\
        \midrule
        LPAE-\textit{u} \cite{LPAE}  & 77.58          & 77.92          & 80.72           & 80.97          \\
        LPAE-\textit{g} \cite{LPAE}  & 77.65          & 77.35          & 82.62           & 82.53          \\
        \midrule
        FND                          & 78.68          & 79.42          & 83.84           & 84.10          \\
        FND (w)                      & 79.72          & 80.98          & 84.53           & 85.31          \\
        FND-CL                       & 79.59          & 80.09          & 85.26           & 85.43          \\
        FND-CL (w)                   & \textbf{80.33} & \textbf{81.56} & \textbf{85.51}  & \textbf{86.41} \\
        \bottomrule
    \end{tabular}
\end{table}

        We evaluate the ranking performance of recommending to the cold starters.
        Concretely, we test a scenario where a model is trained in Polyvore-\{630, 519\}, and cold starters in Polyvore-\{53, 32\} desire recommendations.
        Following the previous work \cite{LPAE}, we experiment on the circumstances that each cold starter has only 1 or only 5 interacted outfits.
        We compare our methods with non-personalized and LPAE methods, which do not require additional training of the model.
        For LPAE methods, we use the anchor-search \cite{LPAE}, which is known to be the most effective strategy in the cold-start case.
        In FND and FND-CL, we evaluate both \textit{avg} and \textit{w-avg} strategies.
        We conduct experiments 10 times and report the average results in Table \ref{tab:cold}.
        The results show that our approaches consistently outperform baseline methods even though the primary purpose of LPAE methods is to deal with the cold-start problem.
        The \textit{w-avg} strategy is more effective than the \textit{avg} strategy in both FND and FND-CL, implying the importance of considering neighbors differently based on similarity rather than treating them equally.

    \subsection{Performance on Hard Negative Outfits}
    \label{sec:hard}

\begin{table}
  \caption{Comparison of different methods on hard negative outfits.}
  \label{tab:hard}
  \begin{tabular}{lcccc}
    \toprule
                                 & \multicolumn{2}{c}{Polyvore-630-H} & \multicolumn{2}{c}{Polyvore-519-H} \\
    \midrule
    Method                       & AUC             & NDCG             & AUC             & NDCG             \\
    \midrule
    \midrule
    OutfitNet \cite{OutfitNet}   & 79.19           & 60.40            & 79.83           & 60.77            \\
    FHN \cite{FHN}               & 79.67           & 60.61            & 80.15           & 60.20            \\
    LPAE-\textit{u} \cite{LPAE}  & 81.69           & 65.26            & 82.17           & 65.79            \\
    LPAE-\textit{g} \cite{LPAE}  & 80.88           & 63.99            & 80.53           & 62.35            \\
    \midrule
    FND                          & 83.55           & 68.74            & 83.90           & 69.11            \\
    FND-CL                       & \textbf{84.24}  & \textbf{69.99}   & \textbf{84.76}  & \textbf{70.50}   \\
    \bottomrule
  \end{tabular}
\end{table}
        
        We test a more challenging case where negative outfits in evaluation are composed of positive outfits of other users (\ie hard negatives).
        We only compare with personalized methods since non-personalized methods cannot distinguish users.
        Following the previous work \cite{FHN}, we set half of the negative outfits to hard negative outfits when training the model.
        We report the results in Table \ref{tab:hard}.
        The results show that LPAE-\textit{g}, which additionally considers non-personalized compatibility, performs poorly in this task compared to LPAE-\textit{u}.
        We can see that our FND and FND-CL outperform the baseline methods.
        Note that the effectiveness of the CL framework is more apparent on hard negative outfits than the results in Table \ref{tab:main}.
        We believe that the functionality of CL to distinguish outfits enables the model to capture more meaningful outfit representations, especially in the hard negative setting.
    
    
    \subsection{Performance with Different $\alpha$}

\begin{figure}[t]
    \centering
    \includegraphics[width=\linewidth]{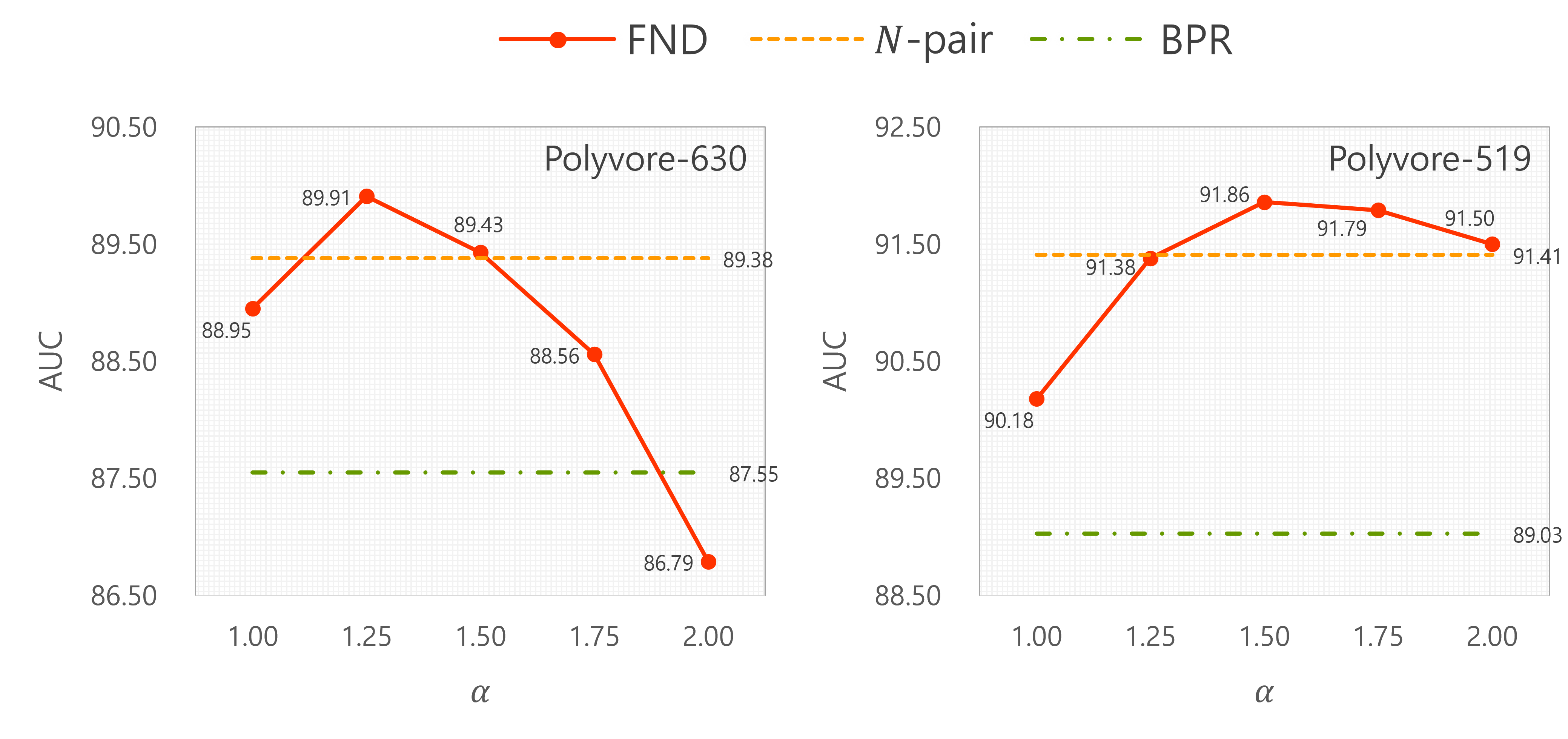}
    \caption{Comparison of different $\alpha$ on Polyvore datasets.}
    \Description{We report the AUC of FND according to alpha on Polyvore datasets. In addition, the performance of N-pair and BPR is also shown in the same figure. For each dataset, FND outperforms other baselines when appropriate alpha is chosen.}
    \label{fig:FND}
\end{figure}

        We evaluate the performance of FND under various $\alpha$ (see Eq. \ref{eq:alpha}).
        The results are reported in Fig. \ref{fig:FND}, and we also show the performance of $N$-pair and BPR in the same figure.
        As mentioned in Sec. \ref{sec:teacher}, $N$-pair overcomes the partial shortcoming of BPR by considering multiple negative outfits in each update and thus clearly outperforms BPR.
        Different datasets tend to have different optimal $\alpha$, but given adequate value, FND can surpass the performance of a strong $N$-pair.


    
    \subsection{Performance with Different Augmentations}

\begin{figure}[t]
    \centering
    \includegraphics[width=\linewidth]{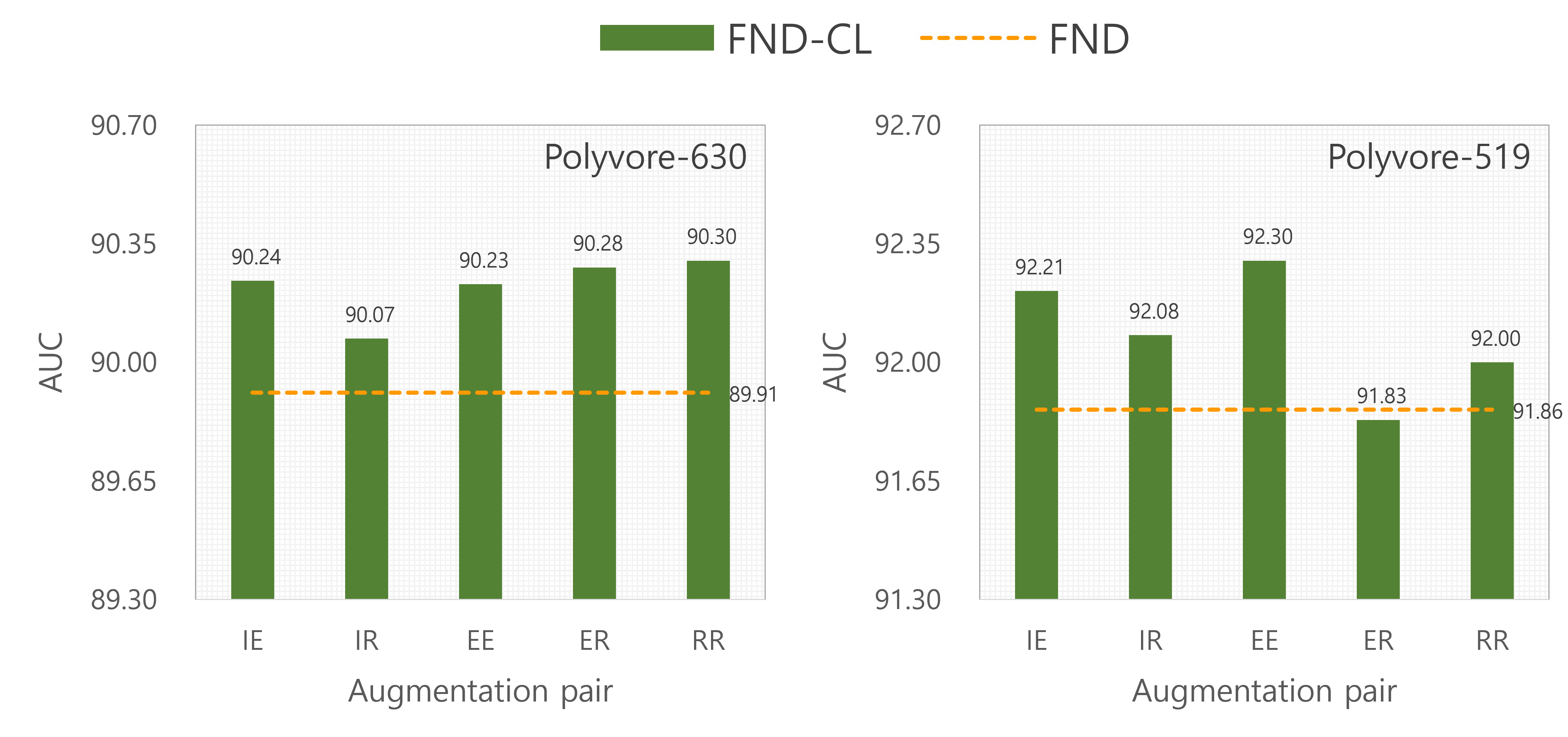}
    \caption{Comparison of different augmentation methods on Polyvore datasets. Augmentation pair XY represents that the first and the second augmentations are X and Y. We also include the identity function as an augmentation method. I/E/R indicates \textit{identity}/\textit{erase}/\textit{replace} augmentations.}
    \Description{We report the AUC of FND-CL according to augmentations on Polyvore datasets. In addition, the performance of FND is also shown in the same figure.}  
    \label{fig:FND-CL}
\end{figure}

        To test the performance of FND-CL for all possible augmentation methods, we put the identity function into the set of augmentations.
        The results are shown in Fig. \ref{fig:FND-CL}, and we also report the performance of FND in the same figure.
        Note that we did not experiment on the (\textit{identity}, \textit{identity}) augmentation pair because two augmented views should be different.
        Regardless of which augmentation pair we use, FND-CL outperforms FND in almost all cases, and the optimal augmentation pair is different for each dataset.
        Concretely, \textit{replace} and \textit{erase} augmentation methods tend to be more effective at Polyvore-630 and Polyvore-519, respectively.
        Therefore, we can see that both \textit{erase} and \textit{replace} are meaningful augmentation methods and that the CL framework is effective in outfit recommendation.

    \subsection{Performance with Different Model Sizes}

\begin{figure}[t]
  \centering
  \includegraphics[width=\linewidth]{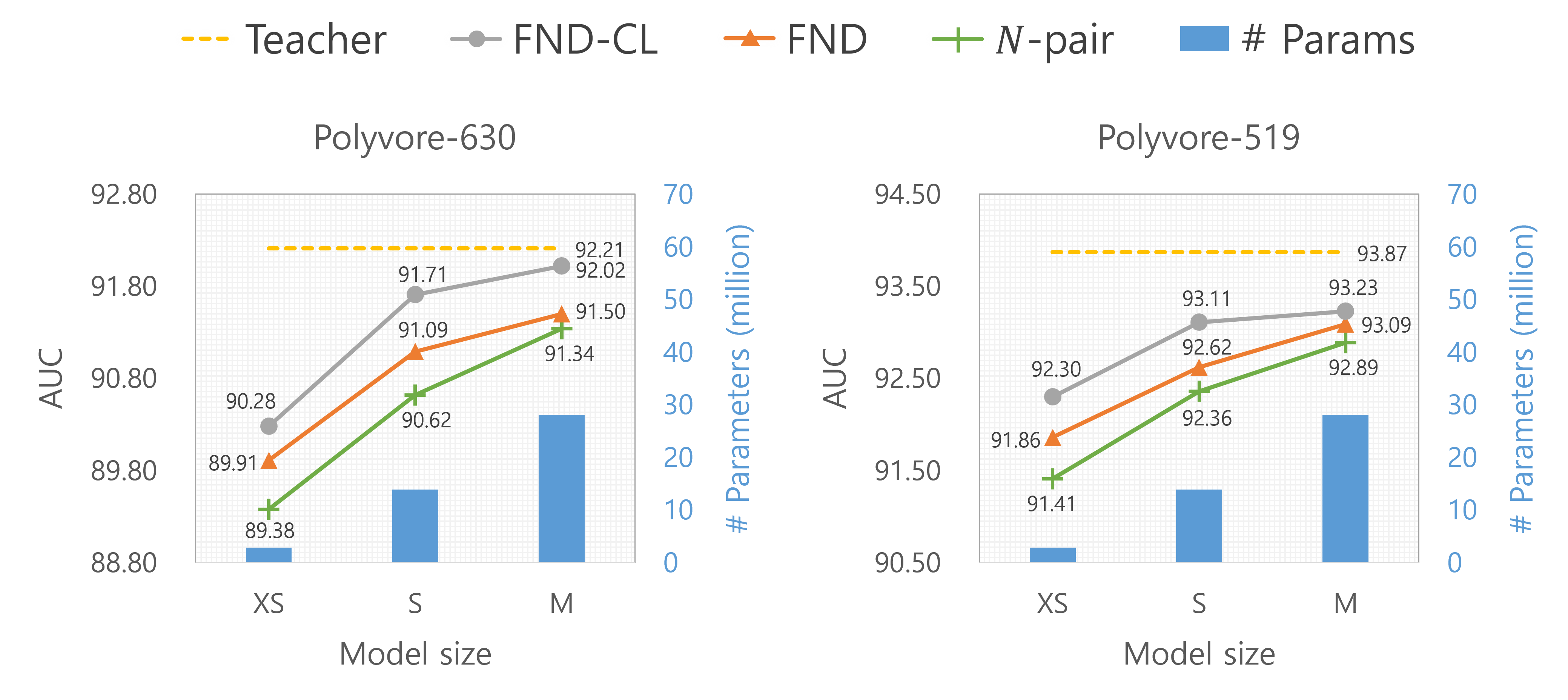}
  \caption{Comparison of different model sizes on Polyvore datasets.}
  \Description{We report the AUC of N-pair, FND, and FND-CL according to model sizes on Polyvore datasets. The performance of the teacher model is also shown in the same figure. We also report the number of parameters for each model size in the same figure.}  
  \label{fig:model size}
\end{figure}

        We study the impact of model size on performance, especially in the case of the student model.
        We consider student models with three different sizes (\ie XS, S, M) and the teacher model.
        XS uses AlexNet-small, and S and M use a downsized version of AlexNet that output dimensions of all fully-connected layers are reduced to 1/4 and 1/2, respectively.
        Fig. \ref{fig:model size} shows the results.
        It is clear that the larger the size, the better the performance.
        Moreover, the fact that FND-CL outperforms FND and FND outperforms $N$-pair is consistent regardless of the size of the model, supporting that our approaches are meaningful.
        
        The performance gap between FND and $N$-pair appears to shrink with the increasing size of the student model.
        Such a tendency implies that the effectiveness of FND depends on the performance gap between the teacher and the student model, which implicitly emphasizes the importance of utilizing the superior teacher model.
        On the other hand, we can see that the effectiveness of the CL framework is hardly affected by the size of the model, as expected.
        
        Detailed information on inference efficiency is measured for each model of different sizes and reported in Table \ref{tab:efficiency}.
        We conduct experiments using FND for the student model and $N$-pair for the teacher model.
        Since FND affects only the training step, FND and $N$-pair share the same inference efficiency.
        Note that if CL is added, only the number of parameters increases by about 0.03M.
        In all inference tests, we use PyTorch with CUDA from Tesla P100 SXM2 GPU and Xeon E5-2690 v4 CPU.
        From the results, we can see that as the size of the model increases, all metrics that indicate inefficiency (\ie Time, Memory, \# Params) also increase.

    \subsection{Performance with Different Batch Sizes}

\begin{table}
  \caption{Model compactness and inference efficiency. ``Time'' denotes model inference time for making recommendations to every user in each dataset, and we report the mean and standard deviation of 10 runs. ``Memory'' represents GPU memory usage. ``Ratio'' indicates the relative parameter size of the student model compared to the teacher model.}
  \label{tab:efficiency}
  \begin{tabular}{llrrrr}
    \toprule
    Dataset                                         & Model   & Time          & Memory & \# Params & Ratio   \\
    \midrule
    \multirow{4}{*}{\shortstack[c]{Polyvore\\-630}} & Teacher & 80.5s$\pm$1.7 & 2.89GB & 65.99M    & -       \\
                                                    & XS      & 74.1s$\pm$1.7 & 2.64GB & 2.85M     & 4.3\%   \\
                                                    & S       & 75.2s$\pm$2.9 & 2.68GB & 13.88M    & 21.0\%  \\
                                                    & M       & 76.3s$\pm$3.3 & 2.75GB & 28.10M    & 42.6\%  \\
    \midrule
    \multirow{4}{*}{\shortstack[c]{Polyvore\\-519}} & Teacher & 65.4s$\pm$2.7 & 3.05GB & 65.97M    & -       \\
                                                    & XS      & 58.3s$\pm$3.0 & 2.80GB & 2.83M     & 4.3\%   \\
                                                    & S       & 61.6s$\pm$2.7 & 2.84GB & 13.86M    & 21.0\%  \\
                                                    & M       & 64.4s$\pm$1.9 & 2.91GB & 28.09M    & 42.6\%  \\
    \bottomrule
  \end{tabular}
\end{table}

\begin{figure}[t]
    \centering
    \includegraphics[width=\linewidth]{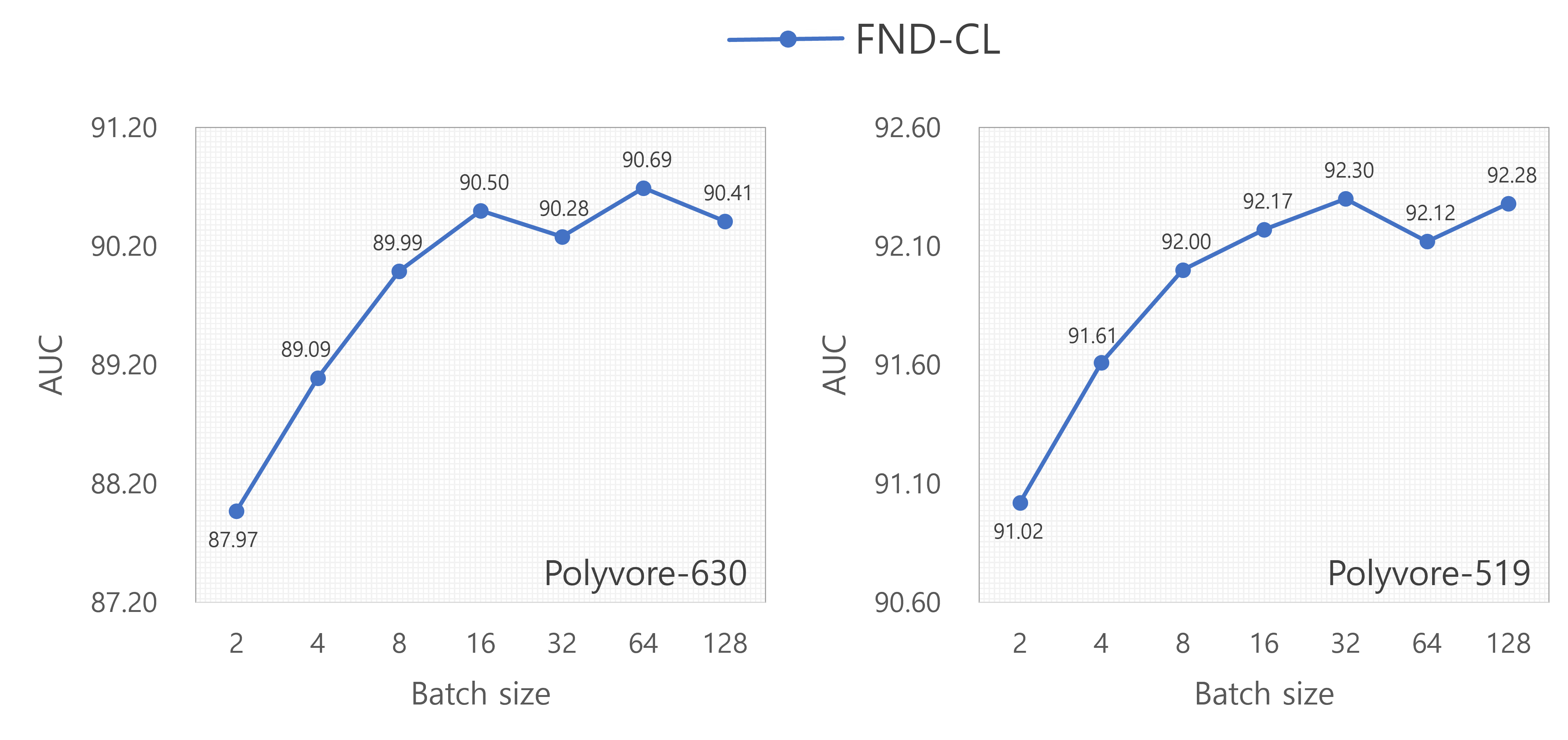}
    \caption{Comparison of different batch sizes on Polyvore datasets.}
    \Description{We report the AUC of FND-CL according to batch sizes on Polyvore datasets.}  
    \label{fig:batch size}
\end{figure}

        To study the impact of the batch size, we test the performance of FND-CL with different batch sizes.
        Note that we use a linear scaling of the learning rate when training with different batch sizes \cite{Krizhevsky2014, Goyal2017}.
        Fig. \ref{fig:batch size} shows the results of the experiment.
        The performance tends to be improved as the batch size increases and appears to converge when the batch size exceeds a certain threshold.
        We believe that the number of negative samples (\ie outfits and augmented views) proportional to the batch size is the primary factor of this tendency.
        Note that except for the number of negative samples, we do not use any other factors significantly influenced by the batch size, such as batch normalization \cite{BatchNorm}.
        Therefore, we can see from the results in Fig. \ref{fig:batch size} as well as Fig. \ref{fig:FND} that it is important to exploit a sufficient number of negative samples per each update when using the ranking loss.

    \subsection{Visualization of the User-Outfit Space}
    
\begin{figure}[t]
    \centering
    \includegraphics[width=\linewidth]{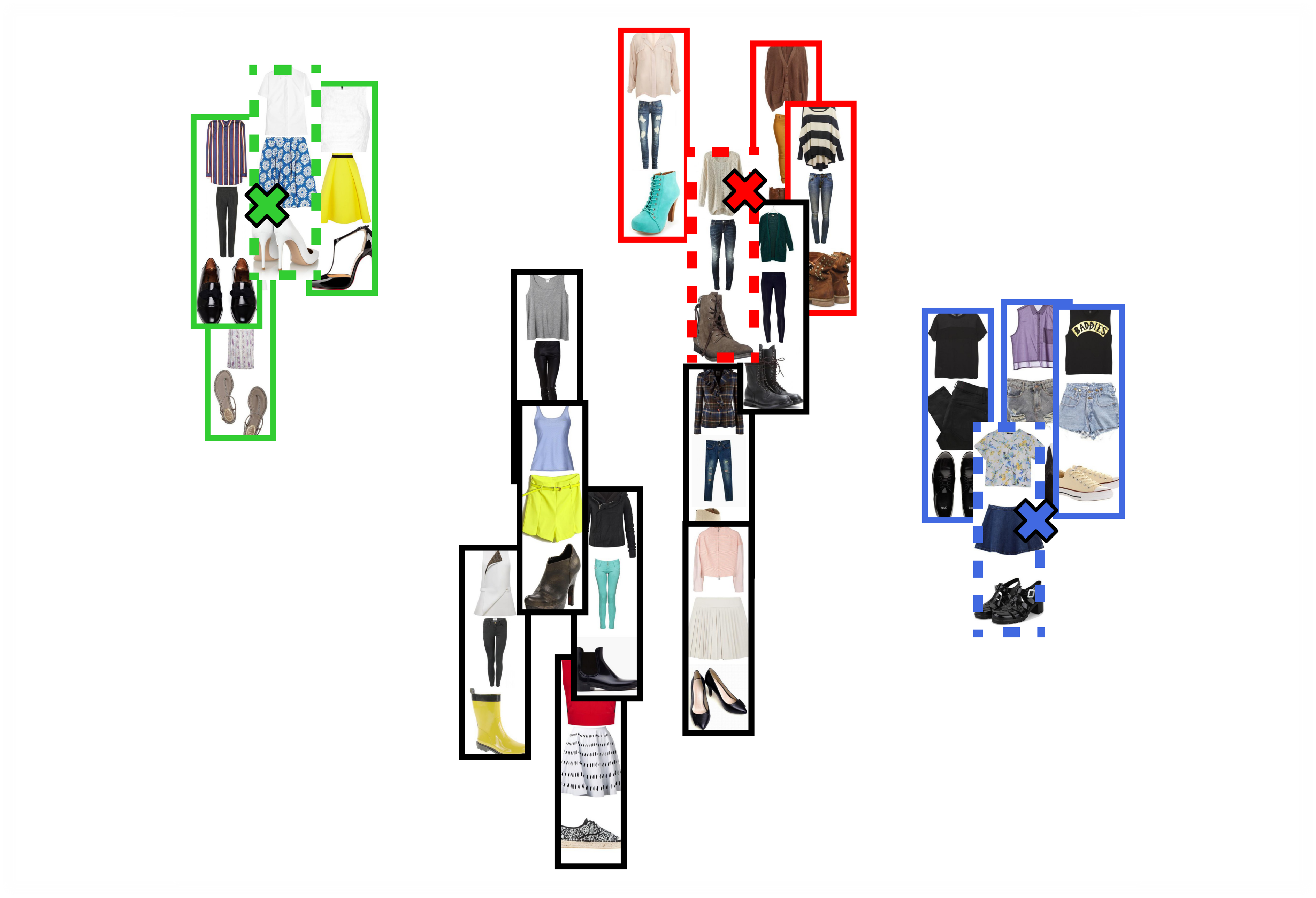}
    \caption{The $t$-SNE visualization result of the user-outfit space. The ``X'' symbol denotes a user embedding vector distinguished by different colors. Each rectangle is an outfit representation vector. A rectangle with a black border indicates a negative outfit. For rectangles with a colored border, they represent the positive outfits of the user corresponding to each color. Among colored rectangles, a dashed border shows a false-negative outfit found in the test set.}
    \Description{We use t-SNE to visualize the user-outfit space of the teacher model on a 2D map. We show users, positive outfits, negative outfits, and false-negative outfits in the figure.}  
    \label{fig:visualization}
\end{figure}

        
        We visualize the user-outfit space of the teacher model to support the intuition of FND (see Fig. \ref{fig:user-outfit space}).
        The visualization uses $t$-SNE \cite{t-SNE} and shows three users and their positive and negative outfits from the training set.
        We focus on the training phase since the approach of FND is to distill knowledge from the teacher model when training the student model.
        The results are shown in Fig. \ref{fig:visualization}.
        Recall that negative outfits are randomly generated, and thus a positive outfit from the test set can appear as a negative sample by pure chance in the training step.
        With the help of the teacher model, the student model can treat such samples as false-negatives, denoted as a dash-bordered rectangle in the figure.
        Moreover, other negative outfits close to a user share a similar style with positive outfits, showing the possibility of being false-negatives.
        Hence, we can conclude that the approach of FND that utilizes false-negative information from the teacher model is reasonable.

\section{Conclusion}

    In this paper, we study how to leverage knowledge distillation (KD) and contrastive learning (CL) framework for personalized outfit recommendation.
    We propose a new KD framework named False Negative Distillation (FND) that does not require the ranking of all possible outfits.
    We also propose two novel data augmentation methods to make use of the CL framework in outfit recommendation.
    Quantitative experiments show that our FND and CL achieve notable success in outfit recommendation tasks.
    In detail, FND outperforms the state-of-the-art methods under fair conditions and achieves improved performance than without using FND in the same model.
    The outfit CL framework also contributes to the recommendation performance by allowing the model to obtain a more meaningful outfit representation.
    We support the soundness of our FND by visualizing the user-outfit space of the teacher model.
    One interesting future work is to apply a contrastive learning framework in a supervised manner by treating each user as a class.



\bibliographystyle{ACM-Reference-Format}
\bibliography{main}

\end{document}